\documentclass[prl,twocolumn,showpacs,preprintnumbers,amsmath,amssymb]{revtex4}

\usepackage{graphicx}
\usepackage{dcolumn}
\usepackage{bm}

\begin{document}

\preprint{APS/123-QED}

\title{Magnetic structure and spin dynamics of \\ quasi-one-dimensional spin-chain antiferromagnet BaCo$_2$V$_2$O$_8$}

\author{Yu Kawasaki$^{1,2}$, Jorge L.~Gavilano$^1$, Lukas Keller$^1$, J\"{u}rg Schefer$^1$, Niels Bech Christensen$^{1,3}$, Alex Amato$^4$, Takashi Ohno$^2$, Yutaka Kishimoto$^2$, Zhangzhen He$^{5,6}$, Yutaka Ueda$^5$, Mitsuru Itoh$^7$}
\affiliation{$^1$Laboratory for Neutron Scattering PSI, CH-5232 Villigen PSI, Switzerland}
\affiliation{$^2$Institute of Technology and Science, The University of Tokushima, Tokushima 770-8506, Japan}
\affiliation{$^3$Materials Research Division, Ris\o{ }National Laboratory for Sustainable Energy, Technical University of Denmark, Frederiksborgvej 399 P.O. Box 49 DK-4000 Roskilde}
\affiliation{$^4$Laboratory for muon Spin Spectroscopy, ETH Z\"urich and PSI, CH-5232 Villigen PSI, Switzerland}
\affiliation{$^5$Institute for Solid State Physics Univ. of Tokyo, Chiba, Japan}
\affiliation{$^6$Fujian Institute of Research on the Structure of Matter, Chinese Academy of Sciences, Fuzhou, Fujian 350002, P. R. China }
\affiliation{$^7$Mat. and Struct. Lab., Tokyo Institute of Technology, Yokohama, Japan}

\date{\today}

\begin{abstract}
We report a neutron diffraction and muon spin relaxation $\mu$SR study of static and dynamical magnetic properties of BaCo$_2$V$_2$O$_8$, a quasi-one-dimensional spin-chain system. A proposed model for the antiferromagnetic structure includes:  a propagation vector ${\vec k}_{AF} = (0, 0, 1)$, independent of external magnetic fields for fields below a critical value $H_c(T)$. The ordered moments, of 2.18 $\mu_B$ per Co ion, are aligned along the crystallographic $c$-axis. Within the  screw chains, along the $c$ axis, the moments are arranged antiferromagnetically. In the basal planes the spins are arranged ferromagnetically (forming zig-zags paths) along one of the axis and antiferromagnetically along the other. The temperature dependence of the sub-lattice magnetization is consistent with the expectations of the 3D Ising model. A similar behavior is observed for the internal static fields at different muon stopping sites. Muon time spectra measured at weak longitudinal fields and temperatures much higher than $T_N$ can be well described using a single muon site with an exponential muon spin relaxation that gradually changes into an stretched exponential on approaching $T_N$. The temperature-induced changes of the relaxation suggest that the Co fluctuations dramatically slow down and the system becomes less homogeneous as it approaches the antiferromagnetic state. 

\end{abstract}

\maketitle
\sloppy

\section{Introduction}

One-dimensional (1D) spin chain systems with nearest-neighbor antiferromagnetic (AF) exchange interactions have attracted much attention in the last decades because they display a rich variety of ground states with non-classical magnetic phenomena.\cite{Tonegawa2002,haldane80,haldane82} It is well known that an ideal 1D AF spin system does not undergo long-range order (LRO) at finite temperatures due to strong quantum spin fluctuations.\cite{bethe31} But slight perturbations  of the system by, for instance,  weak interchain interactions or  doping with non-magnetic impurities, can make the quantum critical state unstable and favors three-dimensional (3D) AF LRO.\cite{GomezSantos90,uchiyama99}

In 2005 BaCo$_2$V$_2$O$_8$, a new member of the quasi-1D AF spin-chain family, was synthesized by Zhangzhen He and co-workers.\cite{he05a,he06a,wichmann86} They succeeded in growing large single crystals of  high quality. BaCo$_2$V$_2$O$_8$ is an insulator with a very anisotropic magnetic susceptibility $\chi$; the easy axis of magnetization is along the crystallographic $c$ axis. The temperature dependence of the magnetic susceptibility $\chi$($T$) displays a broad maximum near 30 K signaling the development of 1D short-range order.\cite{he05a}  

Crystal-field effects and spin-orbit coupling partially lift the degeneracy of the single ion Co$^{2+} (3d^7)$ multiplet resulting in a doublet  ground state. Thus the spin-system may be considered as a collection of effective $S = 1/2$ spin chains.\cite{kimura06} With lowering temperature, a rapid drop of $\chi$ at $T_N = 5.4$ K signals the onset of a long-range ordered AF state at zero field.\cite{he05a} Quantum fluctuations of the spins strongly affect the ordered state, as evidenced by the phenomena observed in external fields applied parallel to the chains, which results in an unusual field-temperature phase diagram (see Fig. 1). 

\begin{figure}[htbp]
\includegraphics[scale=0.5]{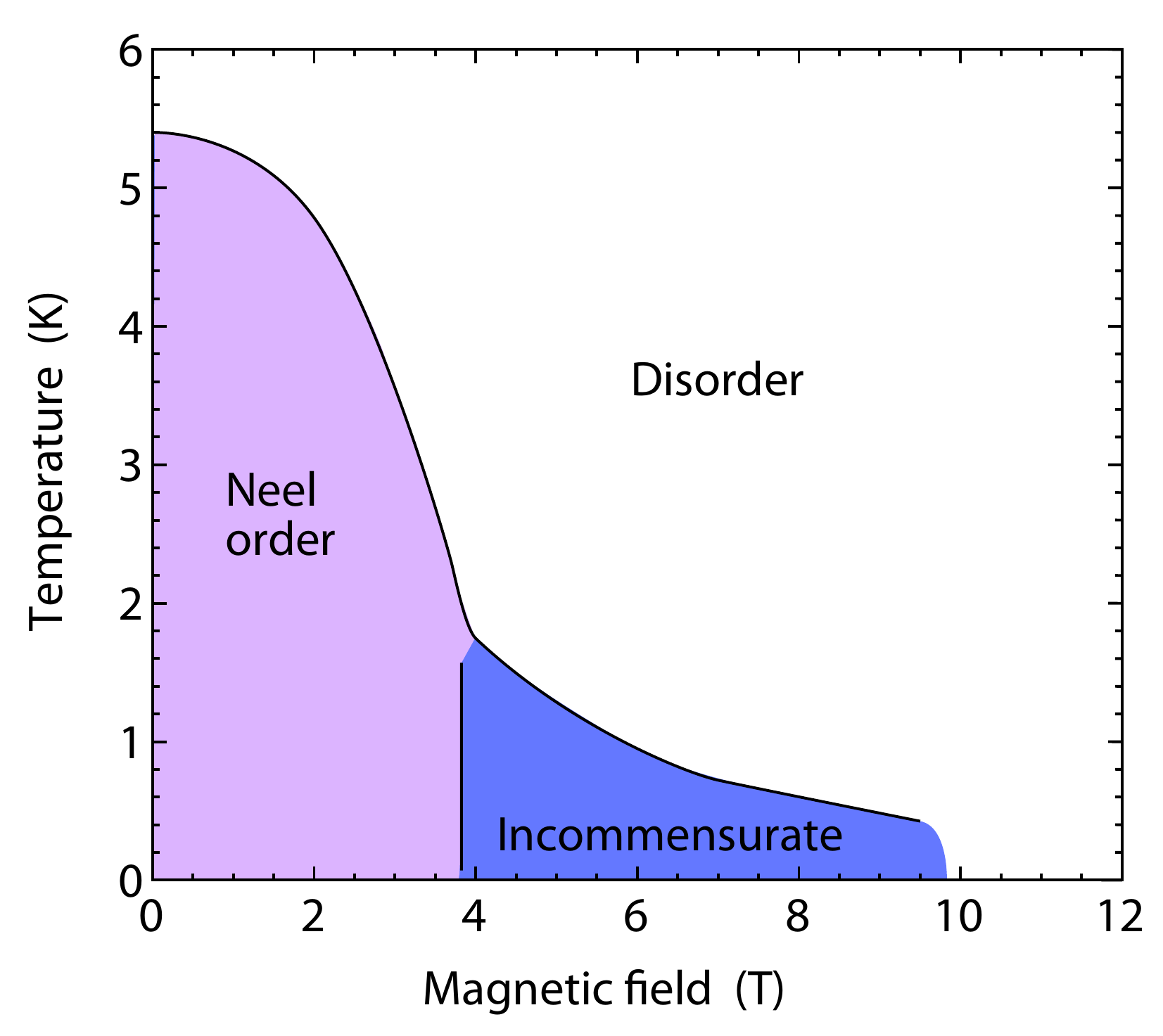}
\caption{Schematic representation of the Magnetic Field - Temperature  phase diagram of BaCo$_2$V$_2$O$_8$ for fields applied along the $c$-axis. Two magnetically ordered phases are distinguished: (i) The N\'{e}el state, a 3D AF state at low fields, and (ii) an incommensurably modulated phase at higher fields (from Kimura and co-workers, Ref. 17). }
\end{figure}

The magnetization curves $M(H)$ are strongly non-linear.\cite{kimura06,kimura07,he06} External magnetic fields $H$ above a critical value $H_c(T)$ suppress the 3D AF N\'{e}el state. At $T = 0$ K, $\mu_0H_c = 3.9$ T. The field-induced suppression of the N\'{e}el  state occurs in an unusual manner. Above 1.8 K, $H$ induces a reentrant phenomenon from the long-range 3D, N\'{e}el  state into a 1D quantum liquid state (order-disorder transition).\cite{he05} This transition was theoretically predicted first for gapped spin systems doped with non-magnetic impurities.\cite{mikeska04} 

At lower temperatures, however,  external fields $ H \ge H_c (T)$ result in the development of an incommensurably modulated AF state. The existence of this novel phase was demonstrated experimentally by Kimura and co-workers.\cite{okunishi07,kimura08,kimura08a,Suga2008} They found for the modulated structure a propagation vector ${\vec k}_M = (0, 0, 1 + \delta) $ with $\delta$  tunable by the external magnetic field. 

Although many interesting properties have been investigated in BaCo$_2$V$_2$O$_8$, there are no reports on the details of the structure of the N\'{e}el ordered state, $i.e.$, in external magnetic fields $H < H_c(T)$. In addition, there is very little information about the spin dynamics in this system, except for two NMR studies.\cite{ideta2010,kuo2010} We  present here a detailed model for the magnetic structure, obtained from neutron diffraction measurements in powder samples and in single crystals of this material. We also report the results of $\mu$SR  measurements below and above $T_N$. The latter shine light into the dynamics of the spin system. 

The physical properties of BaCo$_2$V$_2$O$_8$ can be understood in the context of Ising spin chains with weak interchain interactions as described by the $S=1/2$,  XXZ model where a single chain in the presence of an external magnetic field $H$ is described by the Hamiltonian

\begin{equation}
\mathcal{H} = J  \sum_{i} (S_i^z S_{i+1}^z + \epsilon ( S_i^x S_{i+1}^x + S_i^y S_{i+1}^y ) ) - g\mu_B\sum_{i} S_i^zH
\end{equation}

From magnetization and ESR data it has been determined for BaCo$_2$V$_2$O$_8$  that the AF exchange interaction is $J/k_B = 65$ K, the anisotropy parameter $\epsilon \approx 0.5$, and $g \approx 6.2$.  The interchain interaction $J'$ is of the order of  $J'/J \approx 0.06$.\cite{kimura07} In this model N\'{e}el order occurs at zero field with a non-vanishing gap.\cite{GomezSantos90} The application of  external magnetic fields, $H > H_{c}$, restores the Tomonaga-Luttinger spin-liquid state with longitudinal and transversal incommensurate spin fluctuations that characterize 1D Ising systems.\cite{haldane80, bogoliubov86}

BaCo$_2$V$_2$O$_8$ crystallizes adopting a tetragonal structure with space-group symmetry $I 4 1 /a c d$, room-temperature lattice parameters of $a =  12.444 \AA$ and $c = 8.415 \AA$ and 8 formula units per unit cell. 
All the magnetic Co ions are equivalent. They lie in the centers of CoO$_6$ octahedra, and have strong AF exchange interactions with Co ions in the neighboring  edge-sharing CoO$_6$ octahedra.  The Co$^{2+}$ spins, here referred simply as "spins", form a collection of screw-type chain structures along the $c$ axis. The screw chains are separated by nonmagnetic VO$_4$ (V$^{5+}$) and BaO$_4$ (Ba$^{2+}$)  tetrahedra, resulting in a quasi-one-dimensional structural arrangement of the spins with a strong Ising-type magnetic anisotropy. A representation of the crystal structure is given in Fig. 2.

\begin{figure}[htbp]
\includegraphics[scale=0.8]{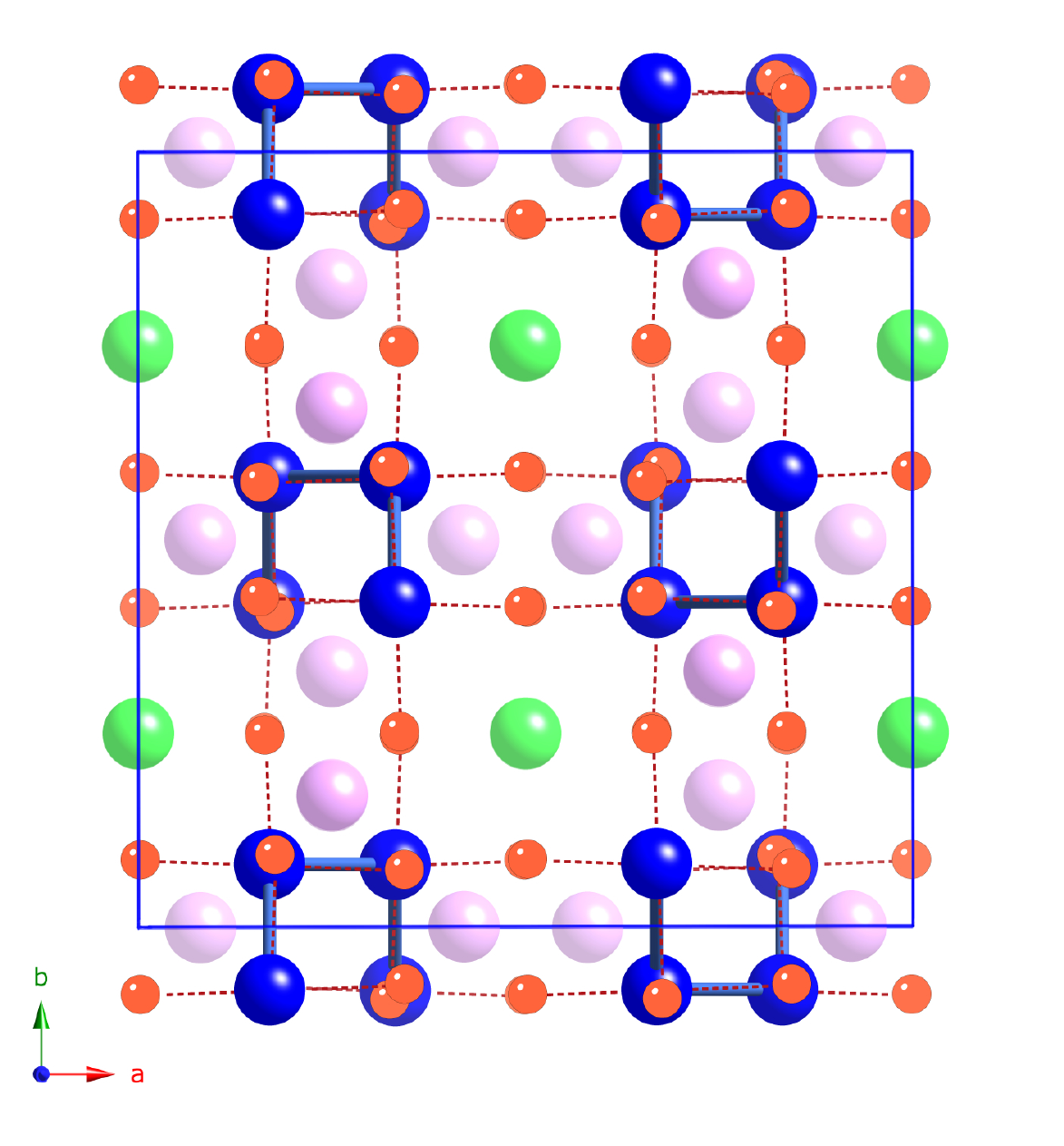}
\caption{Representation of the basal plane projection of  the crystal structure of BaCo$_2$V$_2$O$_8$\@. This figure emphasizes  large blue circles, representing Co ions with the rods joining them representing the n.n.  Co-Co bonds. The small orange circles represent the O ions, the green  and violet circles represent the Ba and V ions, respectively (colors on line). }
\end{figure}

\section{Experimental procedure}

Polycrystalline samples of BaCo$_2$V$_2$O$_8$ were synthesized by a standard solid-state reaction method using high purity reagents of BaCO$_3$ (4N), CoC$_2$O$_4$$\cdot$2H$_2$O (3N) and V$_2$O$_5$ (4N) as starting materials.   Single crystals of BaCo$_2$V$_2$O$_8$ (of dimensions of the order of $10\times8\times3$ mm$^3$) were grown by a spontaneous nucleation method from the same starting materials.  Details of the material synthesis have been previously described.\cite{he05a,he06,he06a} Our material was characterized by X-ray diffraction and magnetic susceptibility. The crystal structure and the magnetic properties confirmed the conclusions based on previous data.\cite{he05}

The magnetic structure of BaCo$_2$V$_2$O$_8$ in zero external magnetic field was investigated by measuring neutron powder diffraction patterns  at temperatures down to 1.8 K using incident neutrons of wavelength of 2.4503 \AA\ in the Cold Neutron Powder Diffractometer DMC at SINQ, Paul Scherrer Institute (PSI), Switzerland. The sample consisted of 8 g of powdered BaCo$_2$V$_2$O$_8$ material enclosed in a thin-walled V cylindrical container.
On a single crystalline sample, we measured the temperature and field dependencies of some selected magnetic Bragg reflections using the Triple Axis Spectrometer Rita-II and the Thermal Single Crystal Diffractometer TriCS both at SINQ, PSI. The latter with incident neutrons of a wavelength $\lambda = 1.18 \AA$.
External magnetic fields, of  up to $\mu_0H =  4.5$ T,  were applied along $c$ axis.
$\mu$SR time spectra were measured on a polycrystalline sample, in the temperature range between 1.5 and 100 K in zero and weak longitudinal magnetic fields, using the General Purpose Surface-Muon Instrument GPS at the Swiss Muon-Source of PSI.

\section{Results and discussion}

\subsection{Magnetic structure}

\begin{figure}[tbp]
\includegraphics[scale=0.5]{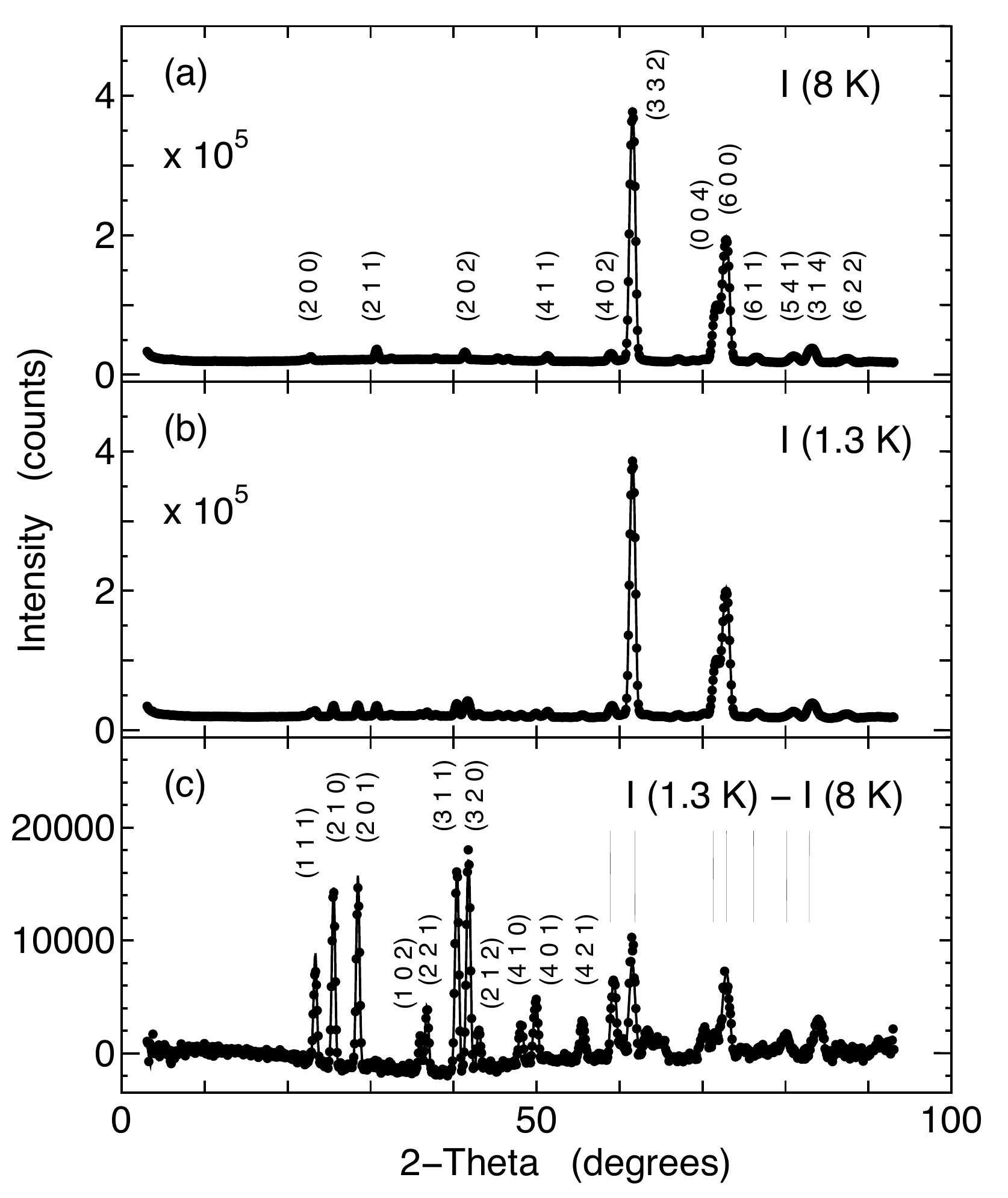}
\caption{Bragg scattering profiles for BaCo$_2$V$_2$O$_8$ at $T$ = 8 K (a) and $T$ = 1.3 K (b). A pattern corresponding to the difference of intensities  [$I$(1.3 K)$-I$(8 K)] and revealing only the magnetic Bragg reflections is shown in (c). The solid lines represent the refinement of the data using a model described in the text.}
\end{figure}

In Figs. 3(a) and 3(b) we display neutron powder diffraction patterns above (8 K) and below (1.3 K) the N\'eel temperature, respectively.  The magnetic Bragg reflections, $i.e.$, the difference of intensities  [$I$(1.3 K)$-I$(8 K)] is displayed in Fig. 3(c). A refinement using the "FullProf" software package \cite{carvajal93} yielded a very good fit of all the magnetic and nonmagnetic reflections. From the nuclear reflections we obtained for the lattice parameters of the tetragonal cell at 8 K: $a$ = 12.389 \AA\ and $c$ = 8.375 \AA.
All the magnetic reflections, which appear below $T_N$, could be indexed in the tetragonal system with a wave vector ${\vec k}_{AF} = (0, 0, 1)$.  The best fit to the data are represented by solid lines in Figs.\ 2(a)-2(c), where the conventional Rietveld factors are $R_{\rm p}$ = 6.08, $R_{\rm wp}$ = 6.13 and $R_{\rm exp}$ = 1.13.

\begin{figure}[htbp]
\includegraphics[scale=0.5]{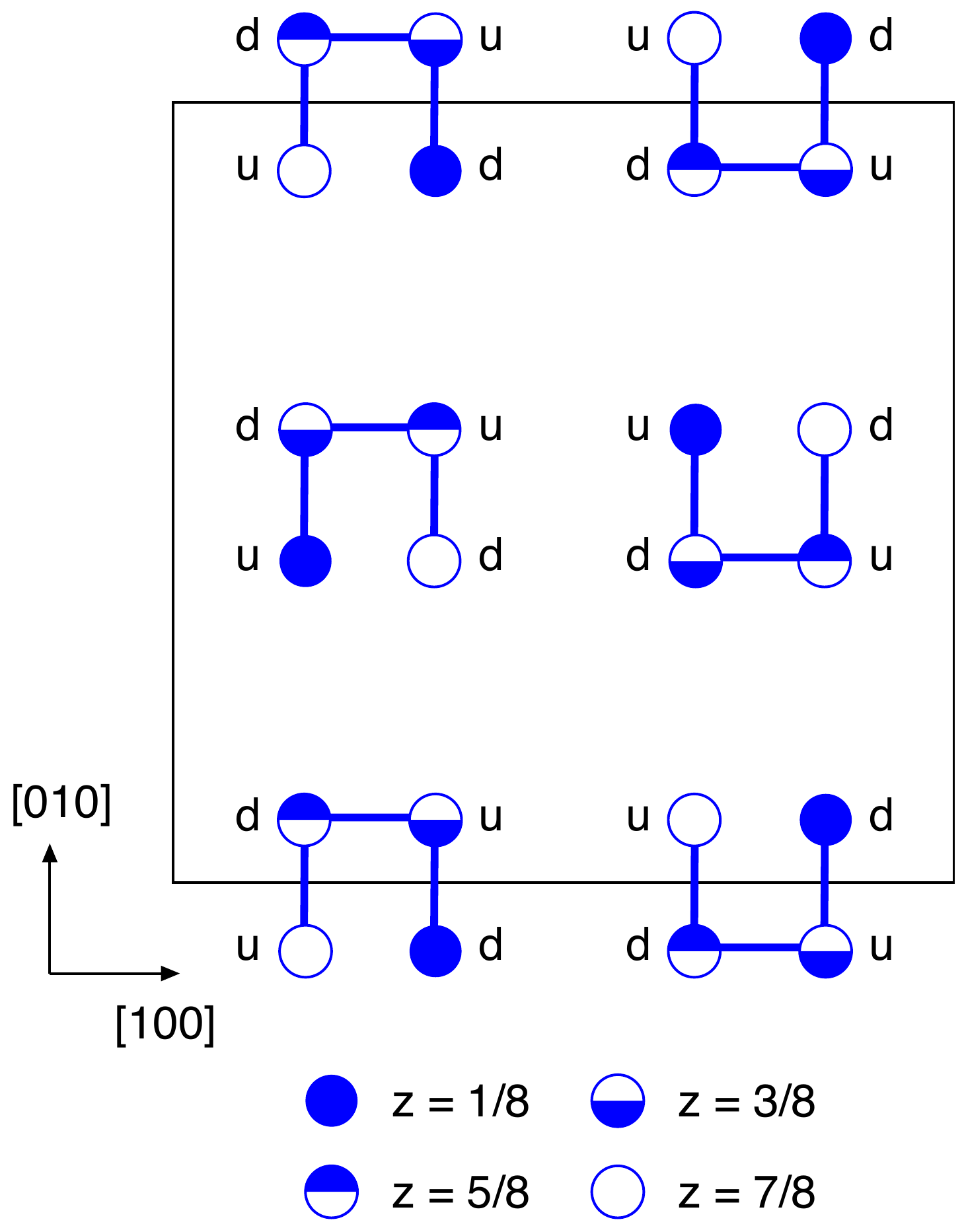}
\caption{Representation of the magnetic structure of BaCo$_2$V$_2$O$_8$ in the AF state. Co ions are depicted as circles and their bondings  as solid lines. The quasi-1D screw-type spin chains run along the $c$-axis and the  "up" and "down" spin directions are represented by ``u'' and ``d'', respectively.}
\end{figure}

The refined spin structure is represented in Fig. 4, where the Co ions are depicted as circles and their bondings  as the heavy solid lines joining  them.  Different coordinates along the $c$ axis are indicated by the different fillings of the circles.  This figure emphasizes the parts of the screw chains contained within the unit cell. The chains run along the $c$-axis, the ordered Co moments are aligned along the same direction  and ``u'' and ``d'' represent up and down spin directions.  The spins are arranged antiferromagnetically along the screw chains, but  at a given basal plane they are arranged antiferromagnetically (ferromagnetically) forming zig-zag paths along the $a$ ($b$) axis. The Bragg peaks are very narrow, resolution limited, thus, the 3D long-range character of the AF order is confirmed.
From the refinement, the magnitude of Co moments is estimated to be 2.18 $\mu_{\rm B}$ which for an effective spin $S=1/2$ implies an effective $g$-factor of $g \approx 4.4$.  

Our neutron powder-diffraction data are not compatible with the value $g = 6.2$, obtained from high-field ESR and magnetization results.\cite{he05a} Assuming the ESR $g$-value, the calculated intensities of the magnetic reflections would be too large, by a factor of 2, to fit the experimental data. Therefore, we conclude that the N\'{e}el state involves reduced Co moments. This reduction is probably due to quantum fluctuations expected in the 3D XXZ model. But the discrepancy of the $g$-factors extracted from neutron scattering, high-field ESR and magnetization results have to be considered carefully since these techniques probe the spin system under different conditions.Ó

\begin{figure}[tbp]
\includegraphics[scale=0.46]{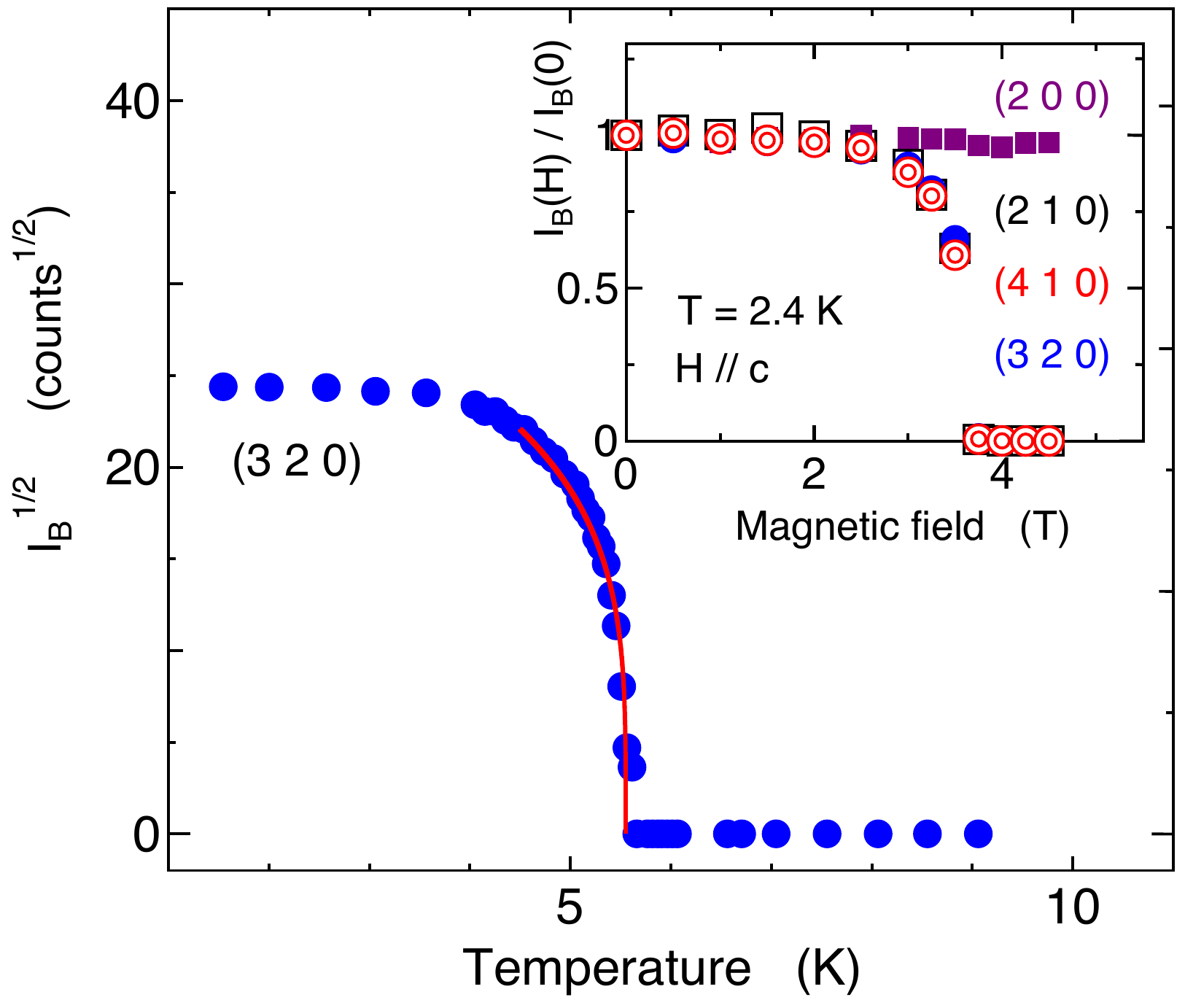}
\caption{The temperature dependence of the square root of the integrated intensity of the (3, 2, 0) magnetic reflection. The inset shows the magnetic field dependence of Bragg intensity of (2, 1, 0), (4, 1, 0), (3, 2, 0) magnetic reflections and the nuclear (2, 0, 0) as a function of magnetic field along the $c$ axis.}
\end{figure}
 
We have measured the temperature and field dependences of selected magnetic Bragg reflections on a single crystalline sample using Rita-II. A summary of these results is presented in Fig. 5.
The circles  in the main panel represent the square root of the integrated magnetic Bragg intensity $\sqrt{I_B}$  of the (3, 2, 0) reflection as a function of temperature. This quantity is proportional to the sublattice magnetization. Its functional form, $i.e.$, the $\sqrt{I_B} (T)$ curve,  follows the  expectations for a 3D Ising antiferromagnet, here represented by the solid line. In a limited temperature range near $T_N$ ($0.8 \leq T/T_N \leq 1$), we find
\begin{equation}
 \sqrt{I_B} \propto (1-T/T_N)^\beta .
 \end{equation}
The fits to the data using Eq. (2)  yielded a N\'eel temperature $T_N$ = 5.4 K, as expected, and a critical exponent $\beta$ = 0.28. This value is close to the predictions of the 3D Ising model, which has been estimated to be $ \beta \approx  5/16$, with the exact value depending slightly on the approximations used. For the 2D case: $\beta =1/8$.\cite{deJongh90,Collins89,Ozeki2007,Fisher67} 

The inset of Fig.\ 5 shows the normalized integrated intensity of three magnetic Bragg reflections (2, 1, 0), (4, 1, 0) and (3, 2, 0) as a function of magnetic field along the $c$ axis.
For comparison, we also plot the intensity of the nuclear Bragg reflection (2, 0, 0).
The data indicate that the magnetic ordering is suppressed by external magnetic fields of 3.75 T along the $c$ axis, at 2.4 K\@.
This is in agreement with the results obtained from measurements of the specific heat in external magnetic fields.\cite{he06a,kimura07}
We found  no measurable field dependence for the magnetic propagation vector $k_{AF}$ in the N\'eel ordered phase.

In an effort to add more light to the nature of the phase transition into the spin liquid phase, we have searched for quasi-elastic scattering  in the vicinity of the above mentioned magnetic Bragg reflections. The searches were performed at temperatures near $T_N$ and zero external magnetic field using the thermal single crystal diffractometer TriCS with a 2D detector, at SINQ, PSI. No clear quasi-elastic signals were detected.

\subsection{Muon spin relaxation and rotation}

\begin{figure}[tbp]
\includegraphics[scale=0.46]{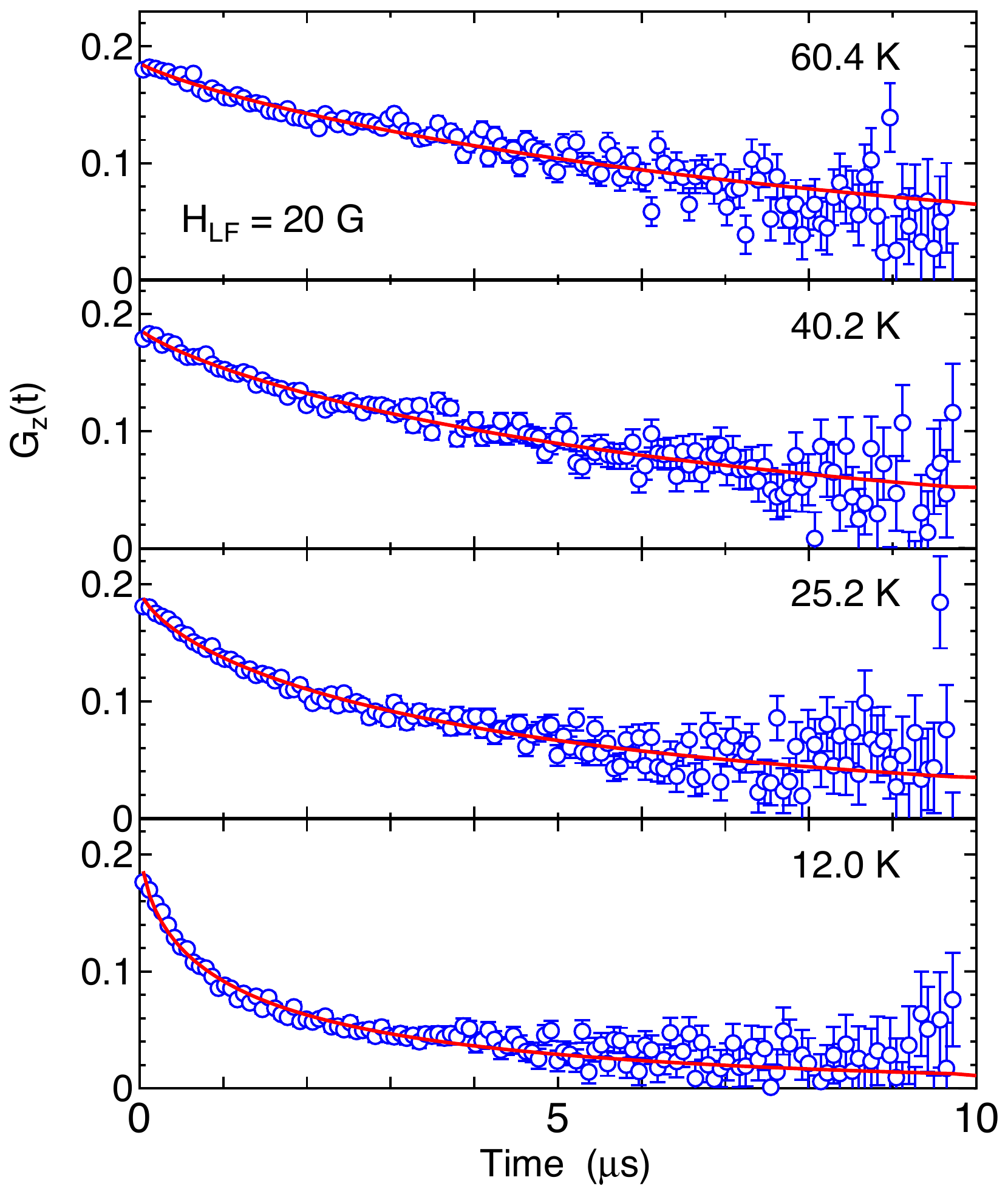}
\caption{$\mu$SR time spectra measured in a weak longitudinal external magnetic field, of 20 G, in the paramagnetic state at four temperatures well above $T_N$\@. The solid lines represent  fits to the data using effectively a single component (see text).}
\end{figure}

To investigate the spin dynamics  of BaCo$_2$V$_2$O$_8$  in the paramagnetic state, we have performed $\mu$SR experiments in zero-field  ZF and in weak longitudinal external magnetic-fields LF.\cite{hayano79}
Four typical $\mu$SR spectra above $T_N$, measured in a field of 20 Oe are shown in Fig.\ 6.
This weak external field effectively suppresses the contribution to the muon relaxation due to quasi-static nuclear dipolar fields. This leaves basically a dynamical relaxation, which  for the case of a single relaxation rate is usually well described by  a simple exponential function.\cite{keren94} More complex dynamic cases are described using stretched exponential functions.\cite{phillips96,phillips95}

\begin{figure}[tbp]
\includegraphics[scale=0.46]{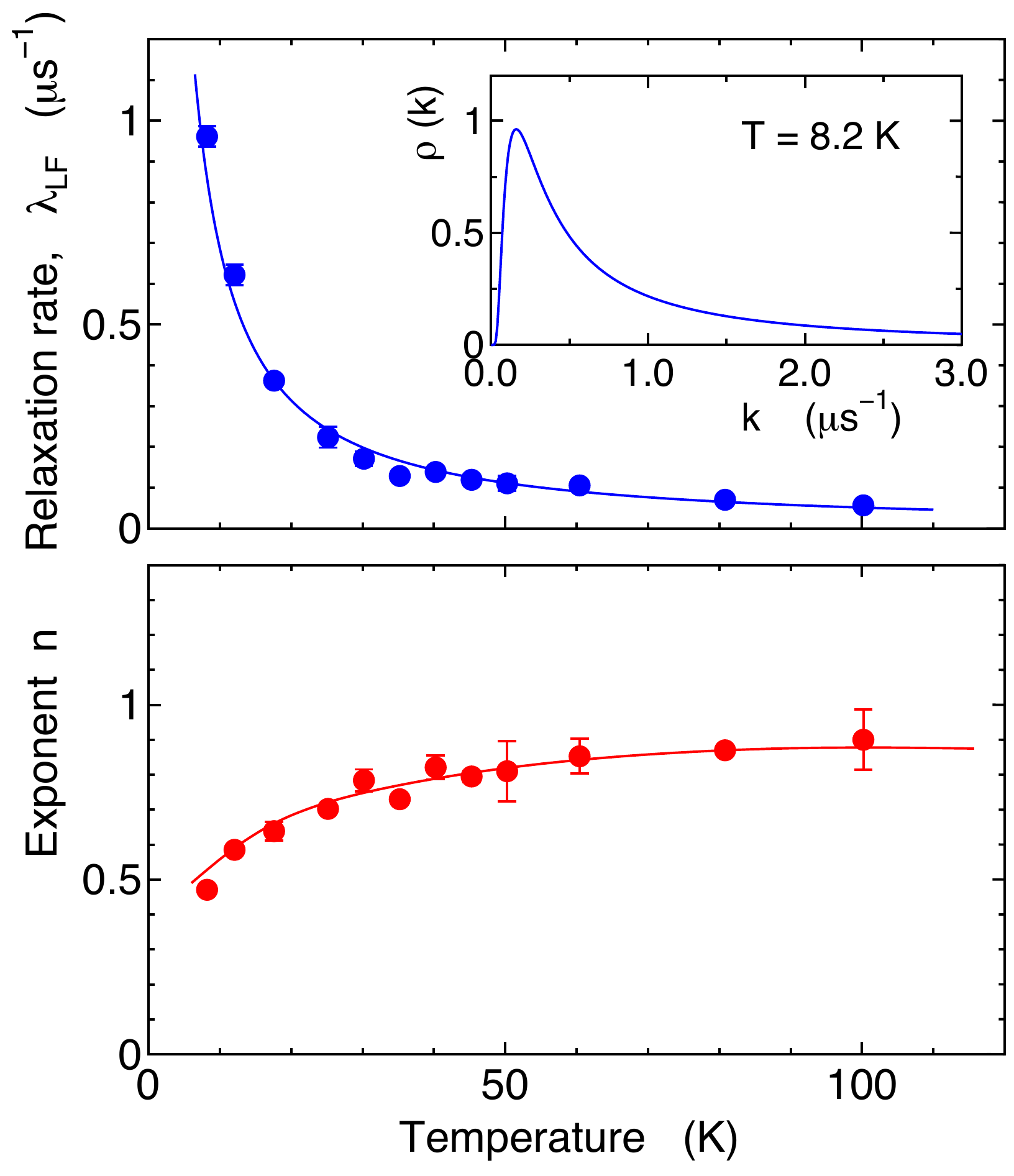}
\caption{Temperature dependence of muon relaxation rate $\lambda_{\rm LF}$ (a) and stretched exponent $n$ (b) obtained in a external longitudinal field of 20 G in the paramagnetic state. The inset in (a) depicts the distribution of relaxation rates at $T \approx 8K$, where $n = 0.5$ (see text). }
\end{figure}

The LF-$\mu$SR time spectra data above $T_N$ were fitted using
\begin{equation}
G_z(t)=A\exp[-(\lambda_{\rm LF} t)^n], 
 \end{equation}
where $A$ is the initial asymmetry  and $\lambda_{\rm LF}$  is a parameter also denoted here as the muon longitudinal spin relaxation rate.
The best fits are represented by the solid lines in Fig.\ 6.
In the paramagnetic state the relaxation (of  dynamic character) is associated with the fluctuations of the Co moments. An stretched exponent  $n \approx 1$ reveals a sharp distribution of rates, which effectively may be represented by a single spin relaxation rate $\lambda_{ LF}$ for the entire spin system, while smaller values, $n < 1$, suggests a distribution of rates which rapidly broadens with decreasing $n$. Only two parameters are needed to fully describe the distribution of rates: The (stretched) exponent $n$ characterizes the width of the distribution and $1/\lambda_{LF}$ is the time needed for the initial asymmetry $A$ to decay to  $1/e$ of its initial value (independent of $n$). 

The results of the temperature dependence of $\lambda_{\rm LF}$ and $n$ are summarized in Figs.\ 7(a) and 7(b).  At high temperatures, $T > 30 K$,  the data are well fitted with an stretched exponent $n \approx 1$, signaling that the system is rather homogeneous. Below 30 K, with decreasing temperature, $\lambda_{\rm LF}(T)$ displays a substantial increase suggesting a rapid slowing down of the dynamics  of the Co spin fluctuations. These are considered as in the "fast fluctuation regime" at all temperatures of these particular measurements, $T > 8$ K. At the same time the system becomes more inhomogeneous, as judged by the rapid decrease of $n$. At 100 K, $n \approx  0.9$  but just above $T_N$,   $n \approx  0.5$.  The increase of $\lambda_{\rm LF}$ with decreasing $T$ is associated with the increase of AF correlations among Co moments within the individual screw chains. This claim obtains support from previous findings based on results for $\chi(T)$. A broad maximum in $\chi(T)$ signals  the onset of intra-chain magnetic correlations.\cite{he06a}  An illustration of the inhomogeneity of the system at low temperatures is given by the distribution of rates  depicted in the inset of Fig. 7(a). 

Stretched exponential relaxation appears in a wide variety of phenomena,\cite{phillips96} including spin glasses and AF systems with frustrated interactions. In such systems,  the relaxation at high temperatures follows an exponential form and, as the temperature decreases (towards the freezing temperature in spin glasses), the functional form changes into a stretched exponential function\cite{phillips95}
\begin{equation}
G_z(t)=A\exp[-(\lambda t)^\frac{d}{d+2}]  .
 \end{equation}
Here the stretched exponent depends on the effective dimensionality $d$. Qualitatively this general picture applies to our system with a reduced dimensionality, since already at 8.2 K, Fig. 7(b) shows that $n \le 0.5$, clearly below the expectations for a 3D case ($n \le 3/5$). There is  a clear trend of decreasing $n$ with decreasing temperature. At low temperatures the tendency seems to be $n \rightarrow 1/3$, $i.e.$, towards the $d = 1$ case. But this trend changes abruptly upon ordering, where the  effective dimensionality of the system suddenly transforms into 3D.

\begin{figure}[tbp]
\includegraphics[scale=0.46]{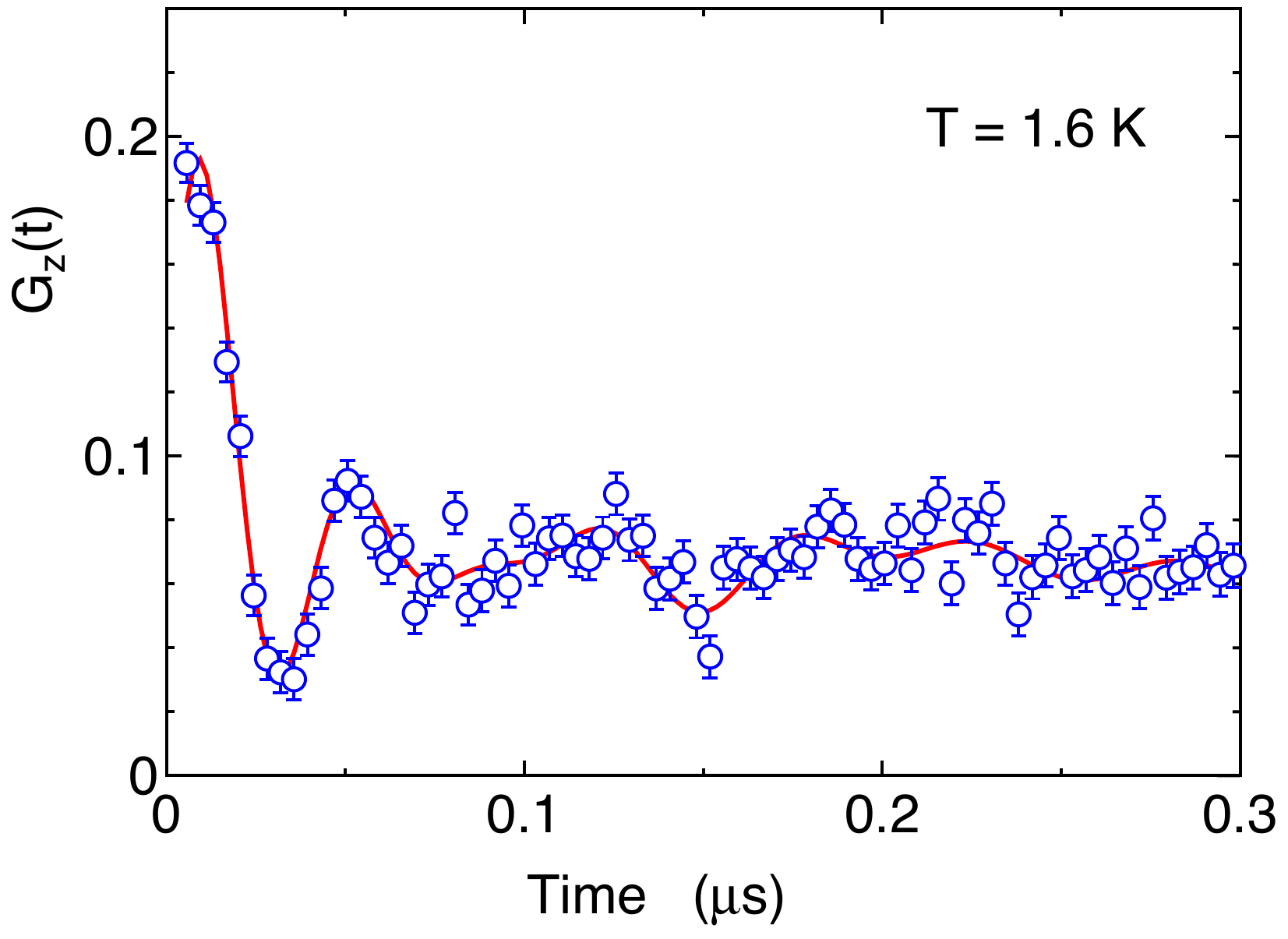}
\caption{$\mu$SR time spectrum measured in zero field in the AF state at 1.6 K\@. The solid line represents a fit to the data using 5 different components.}
\end{figure}

In Fig. 8 we present the ZF-$\mu$SR  time spectrum at $ T = 1.6 $ K, deep in the AF state. Here a spontaneous Larmor precession is revealed by the time-evolution of the asymmetry. The prominent structures revealed in this plot signal the presence of several inequivalent muon sites with non-vanishing internal static fields. This is characteristic  of a magnetically ordered state.  

The muon spectrum below $T_N$ is well described by assuming four components with different relaxation rates and different Larmor frequencies:
\begin{equation}
G_z(t)=A\exp(-\lambda t) +  \sum_{i=1}^{4} A_i\exp(-\lambda_i t)\cos(2\pi\nu_i t + \phi) ,
 \end{equation}
where the first term, a signal with vanishing Larmor frequency, is associated with muons whose spins are oriented along the internal static fields. This corresponds to 1/3 of the total signal for our polycrystalline sample, $i.e.$,  $A \approx  1/3(A + A_1 + A_2 + A_3 + A_4)$. Therefore,  the contribution to the asymmetry from muon sites with vanishing static magnetic dipolar field must be small, otherwise the fraction of $A$  relative to the total asymmetry would be larger.  The signals with non-vanishing Larmor frequencies: $\nu_1 = 5.7$, $\nu_2 = 11.1$, $\nu_3 = 18.4$ and $\nu_4 =  23.9$ MHz, at 1.6 K,  arise from muon stopped at sites with internal static fields of  0.42, 0.82, 1.36 and 1.76 kOe, respectively.  The largest asymmetry is 0.12 corresponding to $\nu_4$   and is substantially larger than the other signals $(A_4 \gg A_1 \sim A_2 \sim A_3)$. Muon motion in the crystal may be ruled out by the lack of appreciable temperature-induced changes in the asymmetries $A_1, A_2, A_3$ and $A_4$ below $T_N$. 

The model described by Eq. (5) is plausible. For instance,  the largest relaxation rate corresponds to the largest internal fields;   $\lambda_4 \approx 28$ MHz. The other three have substantially smaller relaxation rates, of the order of $\lambda_1  \sim \lambda_2 \sim \lambda_3 \sim  7$ MHz. The lowest rate corresponds to the signal with vanishing Larmor frequency $\lambda  <  0.8$ MHz. Although the model yields good  fits of the data it has limitations; namely, it is difficult to identify the muon sites. We found several  interstitial sites,  $(x, y, z)$ with $x/a, y/b, z/c = \{0, 0.5 \}$, where the electric field vanishes. However, at these sites the calculated dipolar magnetic field also vanishes and, therefore, they are unlikely to contribute to the observed asymmetry.  Muon sites where the dipolar magnetic field correspond to the inferred internal static fields are, for instance, near the centers of sides and faces of some BaO$_4$ tetrahedra. However, at these places the electric field does not vanish. We conclude that most likely  the observed signals must arise from muon bound to oxygen ions.

In Fig. 9 we display the temperature dependence of the non-zero muon Larmor frequencies inferred from the zero-field data. All the frequencies vanish as $ T$ approaches $T_N$.  The solid lines represent fits to the data using the critical behavior $\nu = \nu_0(1 -(T/T_N)^\beta)$.  Experimentally one finds $\beta \sim 0.28$, similar to the value obtained in the analysis of the neutron diffraction data, and close to the expectations of the 3D Ising model. Therefore,  below $T_N$ the data consistently suggest that the ordered moments of BaCo$_2$V$_2$O$_8$ may be considered as a 3D Ising system. 

\begin{figure}[tbp]
\includegraphics[scale=0.46]{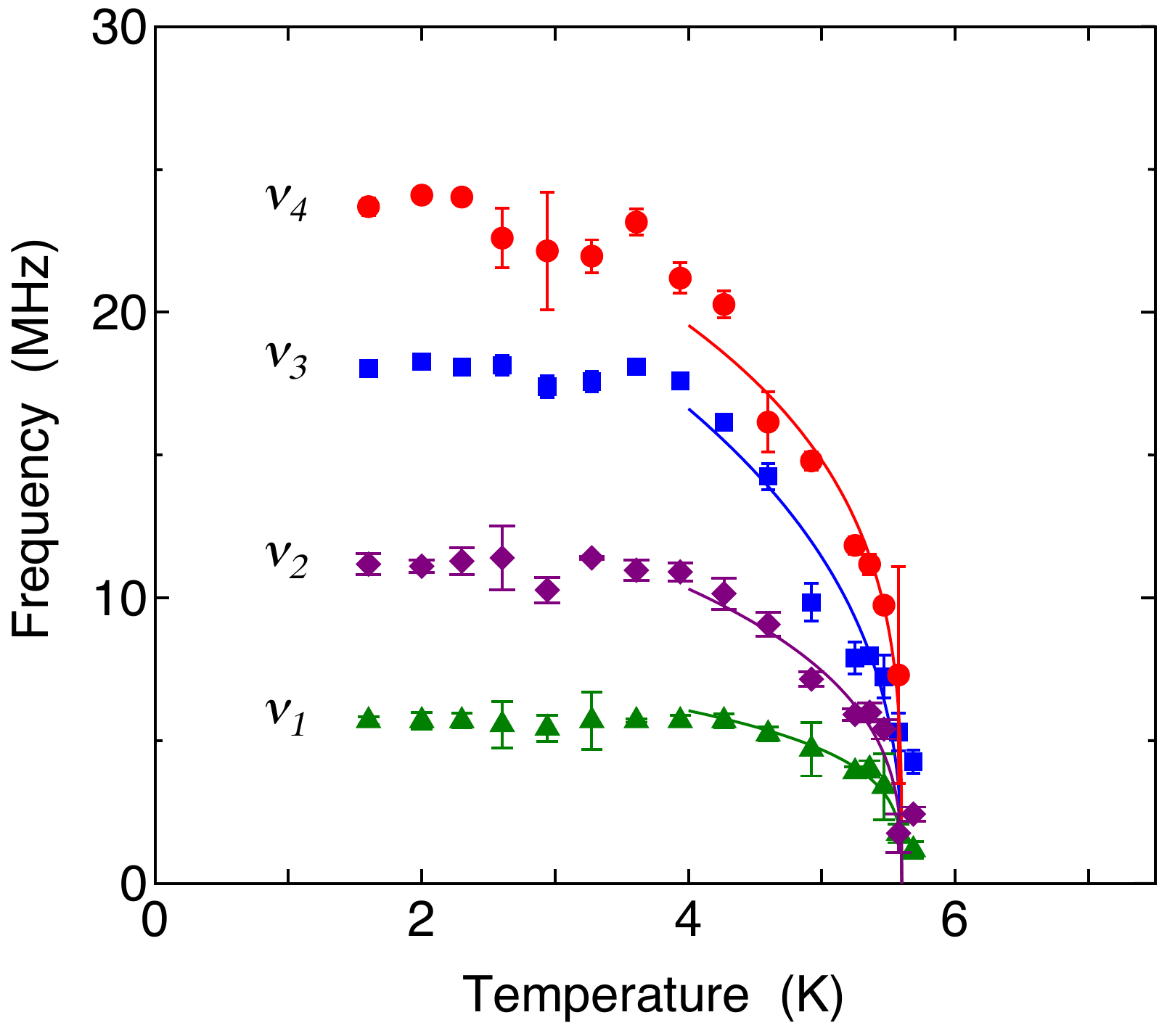}
\caption{Temperature dependence of non-zero muon relaxation frequencies inferred from the zero-field data in the AF state. All the frequencies vanish at $T_N = 5.4$ K. The solid lines represent fits to the data using the 3D Ising model.}
\end{figure}

In summary, as BaCo$_2$V$_2$O$_8$ approaches the antiferromagnetic state  with decreasing temperatures: (i) the Co moment fluctuations undergo a drastic slowing down  and (ii)  the system, as probed by the muon relaxation, becomes gradually inhomogeneous. These features reflect the development of 1D intra-chain AF correlations and show that just above $T_N$ the dynamics of the spin system may be thought of as a disordered Ising  system with reduced dimensionality (probably 1D). The phase transition at $T_N$ may be thought of as the order-disorder transition of 1D objects. Upon order the dimensionality of  system changes into 3. In the AF state, near $T_N$, the temperature dependence of the internal static fields follows the expectations of a 3D Ising model with an effective $S = 1/2$.  The  magnetic structure is characterized by a wave-vector  ${\vec k}_{AF} = (0, 0, 1)$,  independent of temperature and external magnetic field. The moments of 2.18 $\mu_B$ per Co ion are arranged AF within the screw chains, along the $c$ axis, and in the basal planes they form ferromagnetic zig-zag paths along one of the axis and antiferromagnetic paths along the orthogonal (equivalent) direction.

\section{acknowledgments}
This work is based on measurements performed at the laboratory for neutron scattering at the Swiss Neutron Source SINQ  and at the Swiss Muon Source, PSI, Villigen, Switzerland. This work was partially supported by the Danish Natural Science Research Council under DANSCATT (N. C) and by the Advanced International Education and Research Support Program from the Ministry of Education, Culture, Sports, Science and Technology (MEXT) of Japan (Y. K).

\end{document}